\documentstyle [12pt,epsfig]{article}
\begin{document}

\begin {center}
{\Large The kappa in E791 data for $D \to K\pi\pi$}
\vskip 3mm
{D.V.~Bugg}
\vskip 2mm
{Queen Mary, University of London, London E1\,4NS, UK}
\end {center}

\begin{abstract}
A combined fit is made to E791 data on $D \to K\pi \pi$,
LASS data on $K\pi $ elastic scattering, and BES\,II data
on $J/\Psi \to K^*(890)K\pi$.
In all cases, the $K\pi$ S-wave is fitted well with a $\kappa$
pole and $K_0(1430)$; the $\kappa$ requires an $s$-dependent
width with an Adler zero near threshold.
The pole position of the $\kappa$ is at
$\rm {M} = (750 ^{+30}_{-55}) - i(342 \pm 60)$ MeV.
The E791 collaboration fitted their data using a form factor for
the production process $D \to \kappa \pi$.
It is shown that this form factor is not needed. The data
require point-like production with an RMS radius $<0.38$ fm with 95\%
confidence.
\end{abstract}

\vskip 2 mm
PACS: 13.75Cs, 14.40Cs, 14.40.Ev
\newline
Keywords: mesons, resonances
\vskip 2mm

The $\kappa$ in the $K\pi$ S-wave has a controversial history
and is still a confused topic.
There are presently definitive data from three processes.

Firstly, LASS data on $K\pi$ elastic scattering were fitted
originally with an effective range formula and no $\kappa$ pole [1].
This fits the data well.

Secondly,  E791 data on $D^+ \to K^-\pi ^+\pi^+$ revealed a
peak near threshold in the $K\pi$ S-wave [2].
A feature requiring the presence of the
S-wave is strong interference with $K^*(890)$; this  produces a $\cos
\theta$ asymmetry in the $K\pi$ angular distribution.
The threshold peak is isotropic and was fitted with a conventional
Breit-Wigner resonance with mass $M = 797 \pm 19 \pm 43$ MeV,
$\Gamma _{res} = 410 \pm 43 \pm 87$ MeV.
This corresponds to a pole position $\rm {M} - i\Gamma /2 =
(721 \pm 61) - i(291 \pm 131)$ MeV.
The data were later re-analysed by Oller using formulae consistent
with Chiral Perturbation Theory (ChPT) [3]; he reported a $\kappa$
pole at $710 - i310$ MeV.
The CLEO collaboration [4] has reported a Dalitz plot in
$D^0 \to K^0_S\pi ^+\pi ^-$ very similar to that of E791.
The situation is complicated by the appearance of
$\rho (770)$ and other higher $\rho$'s and also the possibility of
$\sigma \to \pi ^+\pi ^-$.
They were able to reproduce their data with a uniform
non-resonant $K\pi$ S-wave; this uniform amplitude has a magnitude
not too far from a broad $\kappa$ resonance.

Thirdly, BES\,II data on $J/\Psi \to K^*(890)K^\pm \pi ^\mp$ exhibit
a low mass $K\pi$ S-wave peak.
It has been fitted with a $\kappa$ pole at $760 \pm 20(stat)
\pm 40 (syst) -i(420 \pm 45(stat) \pm 60(syst))$ MeV [5].
A fit by the BES collaboration to the same data using a
conventional Breit-Wigner parametrisation for the $\kappa$ is given in
Ref. [6].
This fit gives a $\kappa$ pole at $(841 \pm 82) - i(309 \pm 87)$ MeV.

The E791 collaboration has recently refitted
the $K\pi$ S-wave in their data in 37 bins of $K\pi$ mass
over the entire range from threshold to 1707 MeV [7].
In every bin, they determine its magnitude and phase freely,
without prejudice to how it may be interpreted in terms of
$\kappa$, $K_0(1430)$ or any other component.
The objective of the present paper is to show that their data are
accurately consistent with those from LASS and BES if one introduces
into the elastic scattering amplitude the Adler zero of ChPT.
This combined fit eliminates minor inconsistencies between
fits to individual sets of data; it sharpens the conclusions
on the parameters of the $\kappa$.
In passing, it also shows that the $D$ decay
to $\kappa \pi$ is point-like within errors.

The theoretical situation concerning $\sigma$, $\kappa$,
$f_0(980)$ and $a_0(980)$ has been explored
by Oset, Oller, Pelaez and collaborators [8-11].
They take the real parts of
S-wave amplitudes from ChPT and obtain the imaginary parts by
unitarising their amplitudes so as to accomodate rescattering.
The key point is to explain the difference between
elastic scattering and production.
Fig. 1 below shows E791 data, fitted with a variety of form
factors for production.
There is a low mass peak in all cases.
BES data shown in Fig. 4(d) also have a low mass S-wave peak.
However, LASS data on elastic scattering have no peak near
threshold, Fig. 4(b).
Any attempt to fit the production data with the LASS amplitude
is hopelessly bad.

Oset et al. show that the reason for the difference is that
ChPT demands a zero in the elastic scattering amplitude
at the Adler point, $s = s_A = m^2_K - 0.5m^2_\pi \simeq
0.23$ GeV$^2$, not too far from the $K\pi$ threshold
at $s \simeq 0.4$ GeV$^2$.
This Adler zero originates from spontaneous Chiral Symmetry breaking,
which is widely believed to be a key feature of QCD.
The $\kappa$ pole lies in the complex plane with
$Re~s$ close to the $K\pi$ threshold;
the zero distorts the elastic scattering amplitude severely and
goes a long way towards cancelling the nearby $\kappa$
pole.
The zero is not present in production processes, where the
$\kappa $ peak appears undistorted.
They report pole positions for the $\kappa$
of $770 - i(250-425)$ MeV [9] and $750 - i230$ MeV [10].

How may these different situations be accomodated into a
single formula?
The assumption which is tested here is that all data may be
fitted with an $N/D$ form for the  $\kappa$.
The partial wave amplitude for elastic scattering is written
\begin {eqnarray} T_{K\pi \to  K\pi}(s) &=& \frac {N(s)}{D(s)}
  = \frac {\eta \exp(2i\delta) - 1}{2i}, \\
  &=& \frac {g^2_{K\pi  }(s)\rho _{K\pi}}{D_0 - s - i[g^2_{K\pi}(s)
      \rho _{K\pi } + g^2_{K\eta }\rho_{K\eta } + g^2_{K\eta '}\rho
      _{K\eta '}]};
\end {eqnarray}
$\rho (s)$ is Lorentz invariant phase space $2k/\sqrt {s}$
and $k$ is centre of mass momentum.
The Adler zero at $s = s_A$ is included by writing
\begin {equation}
g^2_{K\pi }(s) = \frac {s - s_A}{D_0 - s_A}G
\exp (-\gamma s).
\end {equation}
This is the simplest form which will fit the data.
The exponential cut-off is a detail required to prevent $g^2_{K\pi }$
increasing indefinitely with $s$; BES data require the amplitude to
flatten off or fall at large $s$.
Limitations on the parameter $\gamma$ will be discussed later.
The real part of $D(s)$ is also taken in the simplest form which
will fit the data, $D_0 - s$, with $D_0$ a constant.
An attempt to fit with $D_0$ alone fails to fit the observed phase
variation with $s$.

The denominator $D(s)$ is universal and appears in the amplitude for
production.
However, $N(s)$ is not universal: it depends on the left-hand
cut, which is very different for elastic scattering and production.
Since left-hand singularities for the production process are distant,
the amplitude for production data is taken in the
form
\begin {equation}
T_{D \to \pi \kappa } = \frac {\Lambda \exp (-\alpha q^2)}{D(s)}.
\end {equation}
Here $\Lambda $ is a complex constant
and the exponential is a form factor depending
on the momentum $q$ of the $\kappa$  in the $D$ rest frame.

\begin{figure} [t]
\begin{center}
\epsfig{file=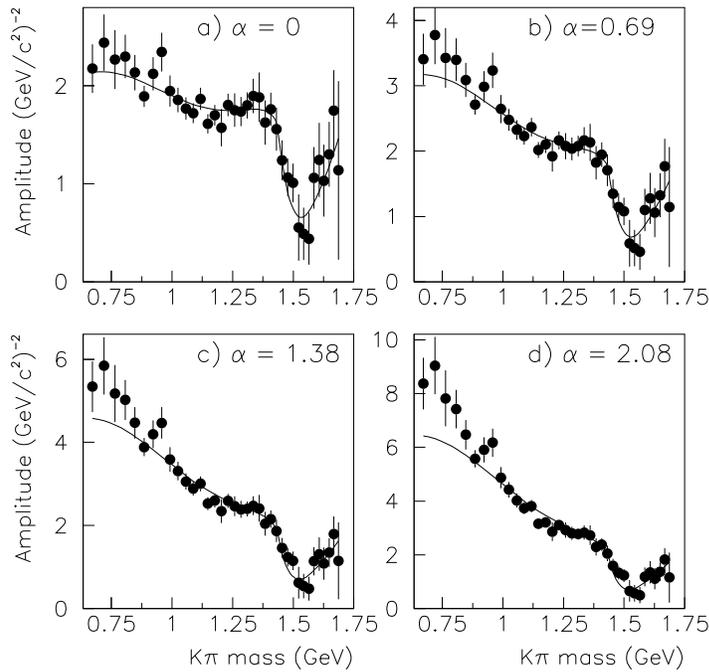,width=10cm} \
\caption{Fits to the magnitude of the $\kappa$ amplitude in
E791 data from Ref. [7] for four values of $\alpha$ in the
form factor; $\alpha$ is in units of GeV$^{-2}$.}
\end{center}
\end{figure}

An important detail is that the Adler zero is a feature of the full
$K\pi$ S-wave.
Therefore it is included into the Breit-Wigner amplitudes for
$K_0(1430)$ and $K_0(1950)$, which are taken in the form of
eqns. (2) and (3).
A detail is that both resonances are consistent with
$g^2_{K\eta } = 0$.
In the LASS data, there are visible resonances at 1430 and 1950 MeV.
Both have tails at low mass.
To keep the amplitude unitary below the inmelastic threshold, the
Dalitz-Tuan prescription [12] is used.
The total S-matrix is written as
the product of $S_i$ for the three individual resonances:
\begin {eqnarray} S_{tot} &=& S_1S_2S_3,   \\
T &=& (S - 1)/(2i),
\end {eqnarray}
For decays of $D$ and $J/\Psi$, there are  large numbers of
open channels, none of which is close to the unitarity limit.
In this case,
unitarity plays no role and the standard isobar-model is adopted,
adding amplitudes from eqn. (4) with fitted complex coupling constants
$\Lambda _i$ for each resonance.

In Ref. [7], E791 adopted the form factor $F = \exp (-\alpha
q^2)$ for the production process $D \to (K\pi )\pi$ with $\alpha = 2.08$
GeV$^{-2}$.
Here this form factor will be adjusted to obtain a fit consistent
with all sets of data.
Fig. 1 shows the fit to E791 data for four values of $\alpha$.
Panel (a) gives the best fit with $\alpha = 0$.
In (b) - (d), the exponent of the form factor is increased in
equal steps to the E791 value in (d).
Fig. 2(a) shows $\chi ^2 $ as a function of $\alpha$.
There is an optimum just below zero for both magnitude and phase,
but consistent with $\alpha = 0$ within the errors.
The RMS radius corresponding to this form factor is zero and
$<0.38$ fm with 95\% confidence.

\begin{figure} [htb]
\begin{center}
\epsfig{file=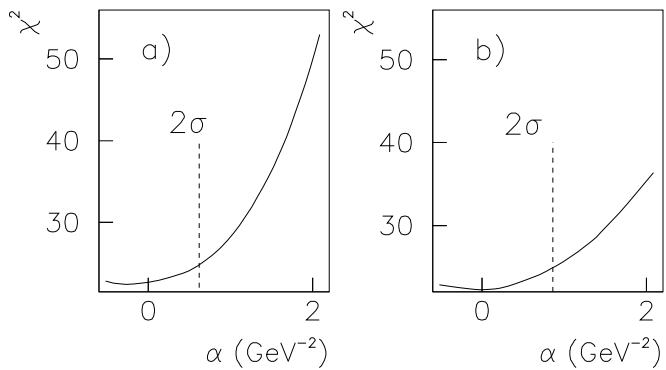,width=10cm}\
\caption{$\chi ^2$ for E791 data as a function of $\alpha$ in the
exponential form factor:
(a) for the magnitude of the amplitude, (b) for the phase.}
\end{center}
\end{figure}

\begin{figure} [t]
\begin{center}
\epsfig{file=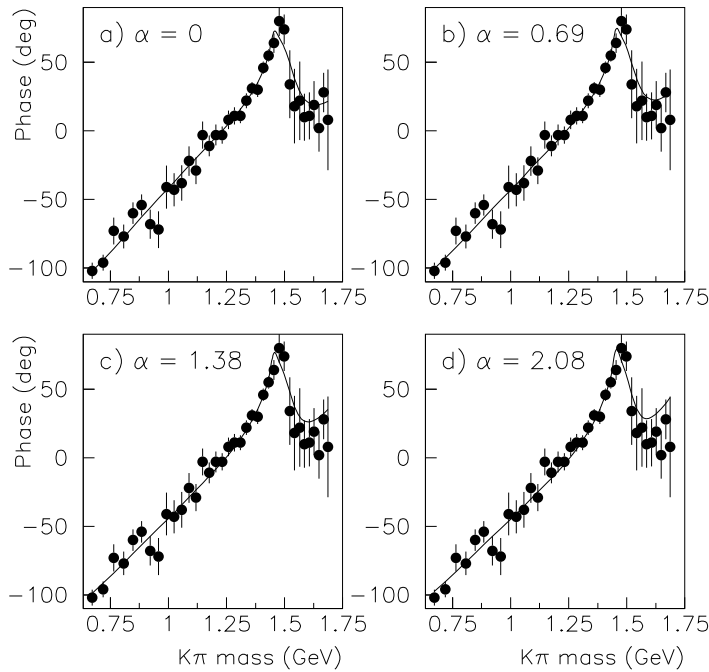,width=10cm}\
\caption{Fits to the phase of the $\kappa$ amplitude in
E791 data for the same $\alpha$ as Fig. 1.}
\end{center}
\end{figure}

Fig. 3 shows corresponding fits to the phase of the $\kappa$.
The sensitivity to $\alpha$ is
less evident to the eye; nonetheless it again optimises
at zero within the errors, as shown on Fig. 2(b).  On careful
inspection of Fig. 3(d), the fit is
systematically high at both ends of the mass range.

A bin-by-bin fit to BES\,II data is reported in Ref. [5].
Magnitudes and phases from that analysis, shown in Figs. 4(c)
and (d), are
fitted here simultaneously with LASS and E791 data.
Variations of $\kappa$ magnitude and phase with $K\pi$ mass are
consistent between LASS and BES\,II data within the errors.

\begin{figure}
\begin{center}
\epsfig{file=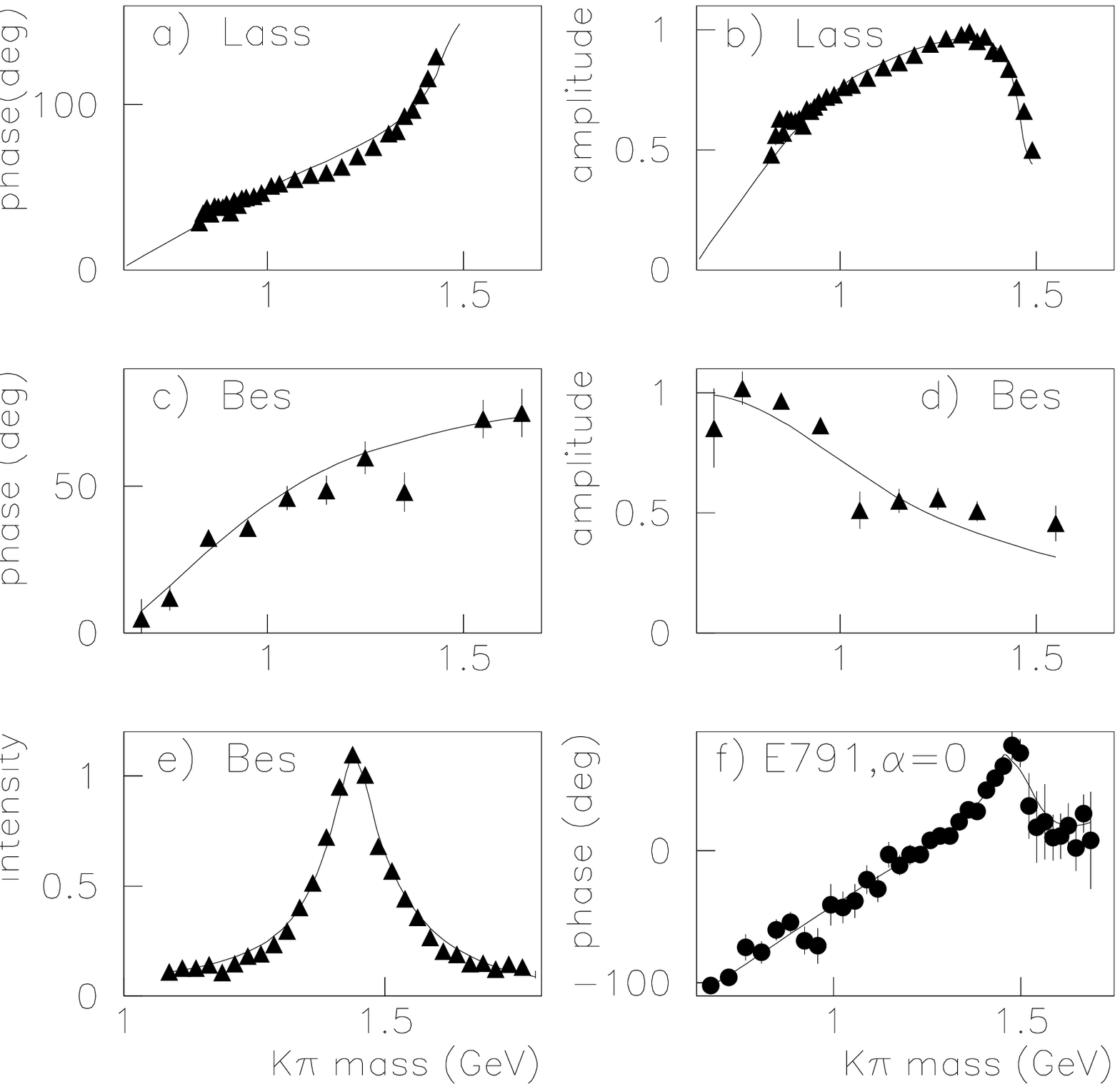,width=14cm}\
\caption{Fits to (a) the phase, (b) the magnitude of LASS
amplitudes for elastic scattering, (c) the phase, (d) the
magnitude of BES data, (e) the 1430 MeV peak  in the $K\pi$
mass projection of BES data, (f) the phase in E791 data.}
\end{center}
\end{figure}

An important ingredient is the mass and width
of $K_0(1430)$.
On  Figs. 1 and 3, it is obviously needed by
E791 data, but a fit to these data alone shows considerable
flexibility in its parameters.
The same is true for LASS data.
In BES data, the $K_0(1430)$ in the channel $K^*(890)K_0(1430)$
is a large signal with well defined centroid and width.
Those data constrain the mass and width of this resonance.
A complication is that there is a smaller
$K_2(1430)$ signal at the same place, but it is well
separated by the amplitude analysis.
Fig. 4(c) shows the mass projection of the 1430 MeV $K\pi$
peak in BES data, allowing an intensity ratio $K_2(1430)/K_0(1430) =
0.75$ [5].

Now we come to numerical details of the fit.
The $K_0(1430)$ amplitude is sensitive to its coupling to
$K\eta '$.
Above this threshold, the
phase of the amplitude varies rapidly with mass.
The opening of  the $K\eta '$ channel also affects the line-shape.
It turns out that E791 data have only weak sensitivity
to the ratio $r = g^2(K\eta ')/g^2(K\pi )$.
If fitted alone, they optimise with zero coupling to $K\eta'$,
but $\chi^2$ changes only by 3.6 for a ratio $r = 1$.
LASS data alone optimise at $r = 1.30 \pm 0.16$.
The full partial wave analysis of BES data is quite
sensitive to $r$ through interferences with other
partial waves and optimises at $ r = 1.0 \pm 0.2$.
We adopt the mean value $r = 1.15$, with an estimated error of
$\pm 0.22$.
The sensitivity of different sets of data to different components
highlights the benefits of fitting several sets of data
together.

Some constraint is needed on the parameter $\gamma $ of
the exponential in  eqn. (4).
It tends to run away to a large value.
However, this leads to a rapid reduction in the width of the $\kappa$
at the top end of the mass range and to an
increase in the $\kappa$ intensity above 1.6 GeV.
That range is not covered by the bin-by-bin data being fitted here, but
is available in the full partial wave analysis.
Those data set upper limits on the magnitude of the $\kappa$ component
between $K\pi$ masses of 1.6 and 2.2 GeV and rule out any significant
increase in the $\kappa$ amplitude in this range.
We therefore constrain the parameter $\alpha$ to be
$\le 0.25$ to avoid such an increase.

The fit to LASS data is not perfect in the mass range 1.15 to
1.35 GeV, see Figs. 4(a) and (b).
It is possible that this is due to some decay to $K\pi \pi \pi$,
presently not known experimentally;
the decay to $\kappa \sigma$ is a possibility which appears
naturally in the linear $\sigma$ model.
Some improvement to the fit is possible by replacing $G$ of eqn. (4)
by $G(1 + \beta s)$.
However, $\beta$ then goes negative and again creates a
problem in the fit to BES data in the mass range 1.6 to 2.2
GeV.
The negative value of $\beta$ produces a zero either in
this mass range or close to it and a large increase in the
$\kappa$ amplitude.
Again this is ruled out by the BES
data.
So we set $\beta = 0$.

The mass and width of $K_0(1950)$ are varied in the
range of values quoted for solution A by LASS.
This solution is favoured by the analysis of
B\" uttiker, Descotes-Genon and Moussallam  [13].
The phases of the highest three $I=1/2$ points of LASS (1.53
GeV upwards) cannot be fitted accurately with  any
variation of parameters and are omitted from the final fit;
these points have little sensitivity to $g^2_{K\eta '}$ or
parameters of $K_0(1950)$.
For the coherent sum of $I=1/2$ and $I=3/2$ amplitudes, LASS
found results lying outside the unitarity circle, so it is
possible that there is a significant sensitivity to the
$I=3/2$ component which was subtracted by LASS.

Equally good fits may be obtained to all data sets for a
large range of $D_0$ in eqns. (2) and (3).
The only real requirement is that $\sqrt {D_0}$ is well above
the mass of $K_0(1430)$, since
the phase of the $\kappa$ rises slowly to $\sim 72^\circ$
at 1.7 GeV.
The value $D_0 = 3.3^2$ GeV$^2$ is adopted from earlier fits
to BES data.
The optimum fit gives $G = 24.5$ GeV$^2$ and $g^2(K\eta )/g^2(K\pi )
= 0.06 \pm 0.02$.
The fit is insensitive to $ r_{\eta '} =
g^2(K\eta ')/g^2(K\pi )$ for the $\kappa$ because of cross-talk with
$K_0(1430) \to K\eta '$.
A range of values for $r_{\eta '}$ is possible from 0 to 0.57 with
little change in $\chi ^2$.
The pole position of the $\kappa$ is rather sensitive to this
variation.
It  varies from $743 - i367$ MeV to $757 - i317$ MeV.
We adopt a mean value $(750 ^{+30}_{-55}) - i(342 \pm 60)$ MeV.
Errors are mostly systematic and depend on variations of the
parameter $\alpha$ of the exponential cut-off for the $\kappa$ and
$r = g^2(K\eta ')/g^2(K\pi ) $ for $K_0(1430)$ and $\kappa$.

For  $K_0(1430)$, fitted parameters are
$M = 1.517$ GeV, $g^2(K\pi ) =  0.353 $ GeV$^2$ for the chosen
value $g^2(K\eta ')/g^2(K\eta ) = 1.15$.
The pole position is at $M = (1406 \pm 29) - i(175 \pm 20)$ MeV.
Data on $K\pi \to K\eta '$ would reduce these systematic errors.
There  is the strong possibility that either or both of $\kappa$ and
$K_0(1430)$ decay to $K\pi \pi \pi$, so data on that channel would
be  helpful.

Some comment is required on the phase of $K\pi$ elastic scattering.
How is it possible to have a pole without the phase going through
$90^\circ$ in the mass range from threshold to 1.4 GeV?
The clue is that the pole lies at $s = 0.445 - i0.513$ GeV$^2$;
its real part lies almost below the $K\pi$ threshold at $s =
0.401$ GeV$^2$.
A Breit-Wigner resonance of constant width would have
a phase of $90^\circ$ at a mass of 667 MeV.
The effect of $\rho (s)$ is to produce a phase rising rapidly from
zero at threshold.
It is essential to realise that the phase varies rapidly as one
goes off the real $s$-axis.
There is a rapid phase variation from the imaginary part of
$(s - s_A)$ and also due to $\rho (s)$ becoming complex.
Near the pole, the phase is $90^\circ$ away from its
threshold value on the real $s$-axis; this was pointed out by Oller in
Ref. [3]. The combined effect of the Adler zero and $\rho (s)$ is to
retard the phase on the real $s$-axis by roughly $90^\circ$ from the
value produced by the pole alone. This is a major distortion from the
familiar resonance of constant width.

Although this may appear surprising, it succeeds in
fitting LASS, E791 and BES data consistently in magnitude and phase.
It is also consistent with ChPT.
Adding one more power of $s$ to either or both of the real and
imaginary parts of $D(s)$ in eqn. (2) does not remove the pole;
the pole position changes within the quoted errors.
The reason for this stability is that LASS and E791 data plus
CHPT determine the phase all the way from the Adler zero to $\sim 1500$
MeV.
Zheng et al. [14] have examined fits to LASS data using a range of
formulae departing from strict ChPT.
Their conclusion is that a pole is required if the scattering length
is $a_0 <0.34 m^{-1}_\pi$.
From the fit reported here, $a_0 = 0.23 \pm 0.04 m^{-1}_\pi$, compared
with the value $0.19m^{-1}_\pi$ of ChPT at order $p^4$[15].

In conclusion, a combined fit to E791, LASS and BES data separates
the $\kappa$ and $K_0(1430) $ components of the $K\pi$ S-wave and
provides a consistent fit to all three sets of data.
They all display the same phase variation as a function of mass
for the $\kappa$ component.
No `background' is required within present errors.
If such background is present, it is the same in all three sets
of data and can be absorbed algebraically into the $s$-dependence
of  the amplitude.
The form factor for production of the $K\pi$ S-wave in
E791 data optimises at $F = 1$, corresponding to point-like production.
The RMS radius for the production process is $<0.38$ fm with 95\%
confidence.

I wish to thank Prof. Brian Meadows for careful and helpful
suggestions.

\end {document}